\documentclass[%
reprint,
 floatfix,
 amsmath,amssymb,
 aps,prl,
]
{revtex4-1}

\usepackage{graphicx}
\usepackage{dcolumn}
\usepackage{bm}
\usepackage[colorlinks=true, citecolor=red, linkcolor=blue,urlcolor=blue ]{hyperref}

\usepackage{lmodern,pdftexcmds,stmaryrd,siunitx,xcolor,tabularx,booktabs,ragged2e}
\DeclareMathAlphabet{\mathsfbi}{OT1}{\sfdefault}{bx}{sl}
\DeclareMathVersion{sfletters}
\SetSymbolFont{letters}{sfletters}{OML}{ntxsfmi}{b}{it}

\newcommand{\bcdot}{\bm{\cdot}}
\newcommand{\bnabla}{\bm{\nabla}}
\newcommand{\btimes}{\bm{\times}}

\newcolumntype{Y}{>{\centering\arraybackslash}X}

\begin{document}

\title{Active particles powered by Quincke rotation in a bulk fluid}

\author{Debasish Das}
\email{dd496@damtp.cam.ac.uk}
\author{Eric Lauga}%
 \email{e.lauga@damtp.cam.ac.uk}
\affiliation{%
Department of Applied Mathematics and Theoretical Physics, Centre for Mathematical Sciences, University of Cambridge, Wilberforce Road, Cambridge CB3 0WA, UK.}%

\date{\today}
\begin{abstract}
Dielectric particles suspended in a weakly conducting fluid are known to spontaneously start rotating under the action of a sufficiently strong uniform DC electric field due to the Quincke rotation instability. This rotation can be converted into translation when the particles are placed near a surface providing useful model systems for active matter. Using a combination of numerical simulations and theoretical models, we demonstrate that it is possible to convert this spontaneous Quincke rotation into spontaneous translation in a plane perpendicular to the electric field in the absence of surfaces by relying on geometrical asymmetry instead.

\end{abstract}

\pacs{Valid PACS appear here}

\maketitle

How are groups of living organisms such as flocks of birds, schools of fish and bacterial colonies  able to self-organize  and display collective motion~\citep{vicsek2012}? This question has fascinated scientists for decades and has given rise to the new field of `active matter'~\citep{marchetti2013,saintillan2018}. 
One of the key features of active matter is that it is composed of self-propelled units that move by consuming energy from their surrounding with a direction of self-propulsion   typically set by their own anisotropy, either in shape or functionalisation, rather than   by an external field.

The origin of macroscopic ordered motion in active systems is due to microscopic interactions occurring at an individual level. Ideally, one would like to develop a coarse-grained description of active systems from these microscopic interactions but  these  are  difficult to measure or quantify, forcing scientists to develop phenomenological models~\citep{vicsek1995,toner1998}. `Non-living' active systems offer a simplified and more controlled setting compared to `living' active systems and there  have been multiple attempts to design self-propelled synthetic particles in the laboratory~\citep{bechinger2016}. Examples include bimetallic Janus particles powered by catalytic reactions~\citep{paxton04,howse2007}, electric~\citep{bricard2013,bricard2015} and magnetic field driven colloids~\citep{kaiser2017}, light activated colloidal surfers~\citep{palacci2013}, water droplets driven by Marangoni stress~\citep{izri2014}, and self-propelled squirming droplets~\citep{thutupalli2011}.

In recent active matter experiments, it has been possible to measure and quantify these microscopic interactions~\citep{bricard2013,bricard2015}. These experiments consisted of spherical colloids able to roll along surfaces by exploiting the so-called  Quincke rotation, discovered more than a century ago~\citep{quincke1896}. The Quincke phenomenon involves the application of a uniform electric field that gives rise to the spontaneous rotation of dielectric solid particles or deformable drops suspended in a slightly conducting fluid medium~\citep{salipante2010,das2013,das2017a}. Quincke rotation is best explained using the much celebrated Melcher--Taylor leaky dielectric model~\citep{melcher1969} that proposes the formation of a surface charge on the particle-liquid interface. Rotation occurs due to the symmetry breaking of the charge distribution that gives rise to a net torque leading to steady rotation of the particle. 

There are two conditions for Quincke rotation to occur. First, the charge relaxation time of the particle, $\tau^-$, must exceed that of the surrounding fluid, $\tau^+$,   
where $\tau^{\pm} = \epsilon^{\pm}/\sigma^{\pm}$ with $\epsilon^{\pm}$ and $\sigma^{\pm}$ being the permittivity and conductivity, respectively (superscript $-$ representing particle and $+$ representing fluid). This implies that the particle must be less conducting than the surrounding fluid, giving rise to a dipole moment, $\bm{P}$, which is anti-parallel to the applied electric field, $\bm{E}_0$. This configuration is unstable and the electric torque, $\bm{T}_E \propto \bm{P} \times \bm{E}_0$, tends to rotate the particle away from its original orientation. The second condition requires that the magnitude of the electric field exceeds a certain critical value, $E_C$, for sustained rotation of the particle, $E_0>E_C$, such that the electric torque  balances the viscous torque.

In an infinite fluid medium, a symmetric particle such as a sphere under Quincke rotation will steadily  rotate  without translating as no net external force acts on it. This spontaneous rotation can be converted into spontaneous translation when the particle is placed near a wall. Such `Quincke rollers' were demonstrated experimentally to perform collective motion due to electrohydrodynamic interactions with each other and with the nearby surface~\citep{bricard2013,bricard2015}.
 
In this Letter, we show that it is possible to convert spontaneous Quincke rotation into spontaneous translation in the absence of surfaces. Specifically, asymmetrically-shaped dielectric particles placed in the bulk of a slightly conducting fluid will spontaneously acquire both rotation and translation under the action of a sufficiently strong uniform DC electric field in a plane perpendicular to the field. We demonstrate this phenomenon by focusing on the electrohydrodynamics of a helix -- an archetypal chiral particle -- first computationally, using the boundary element method, and then by developing an analytical theory in quantitative agreement with the simulations.

\begin{figure}
    \centering
    \includegraphics[width=0.95\linewidth]{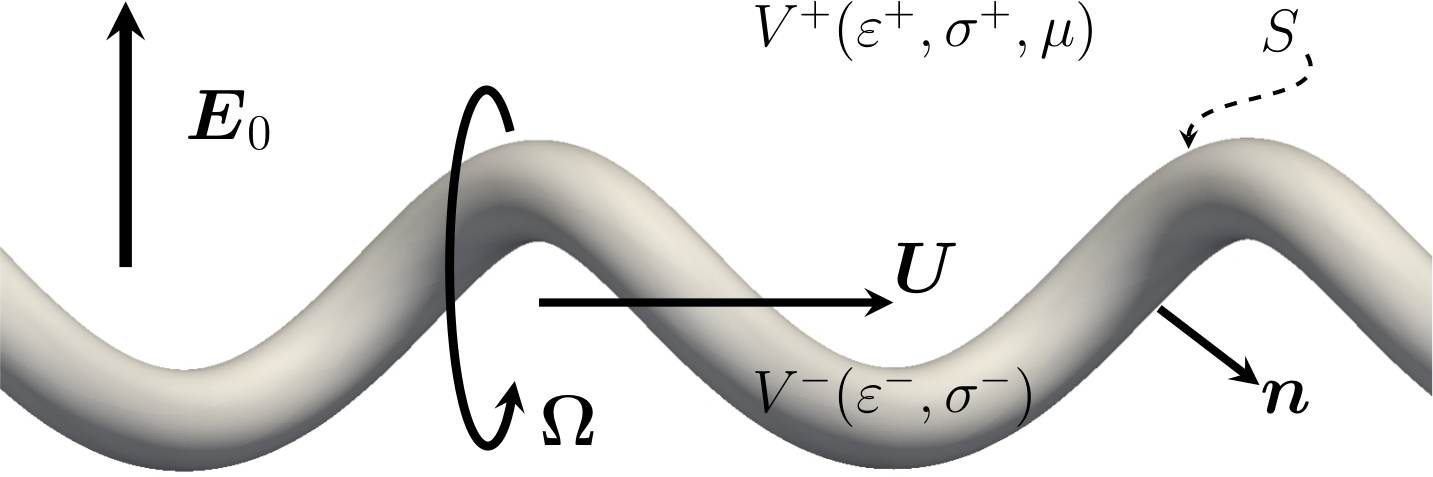}
    \caption{Schematic representation of the problem considered in this Letter.  A solid particle of volume $V^{-}$, surface $S$ and outward unit normal vector $\bm{n}$ is suspended in an infinite fluid  of volume $V^{+}$ and subject to a uniform DC electric field, $\bm{E}_{0} = E_0 \hat{\bm{z}}$. The electric permittivities and conductivities of the suspending fluid and particle are denoted as $\varepsilon^+,\sigma^+$ and $\varepsilon^-,\sigma^-$, respectively, and the dynamic viscosity of the fluid is $\mu$. The translational and angular velocity of the particle are $\bm{U}$ and $\bm{\Omega}$, respectively.}
    \label{fig1}
\end{figure}

Consider an uncharged neutrally buoyant solid particle of volume, $V^{-}$, surface, $S$, and outward unit normal vector, $\bm{n}$, suspended in an infinite fluid medium of volume, $V^{+}$ (see Fig.~\ref{fig1}). The dynamic viscosity of the fluid is denoted by $\mu$. The particle gets polarised due to the application of a uniform DC electric field, $\bm{E}_{0} = E_0 \hat{\bm{z}}$. We define two dimensionless numbers $R=\sigma^+/\sigma^-$ and $Q= \varepsilon^-/\varepsilon^+$ such that $RQ = \tau^-/\tau^+ > 1$ is the necessary condition for Quincke rotation to take place.  In the Melcher--Taylor leaky dielectric model, all charges   are concentrated on the particle surface, so that the electric potential in each domain satisfies Laplace's equation $\nabla^{2}\varphi^{\pm}=0$~\citep{melcher1969}. All the physical quantities are implicitly assumed to be a function of time. On the particle surface, the electric potential and the tangential component of the local electric field are continuous $\llbracket \varphi (\bm{x})\rrbracket =0$ and $\llbracket \bm{E}_t(\bm{x})\rrbracket =\bm{0}$ for $\bm{x}\in S$, where $\bm{E}^{\pm}_{t}=({\mathsfbi{I}}-\bm{nn})\bcdot \bm{E}^{\pm}$, $\bm{E}^{\pm}=-\bnabla \varphi^{\pm}$ and $\llbracket f(\bm{x})\rrbracket \equiv f^+(\bm{x}) - f^-(\bm{x})$ denotes the jump for any field variable $f(\bm{x})$ defined on both sides of the particle surface. The normal component of the electric field $E_{n}^{\pm}=\bm{n}\bcdot\bm{E}^{\pm}$ undergoes a jump due to the mismatch in electrical properties between the two media~\citep{landau1984}, resulting in a surface charge distribution given by Gauss's law, $q(\bm{x})=\llbracket \varepsilon {E}_n(\bm{x})\rrbracket$ for $\bm{x} \in S$.  The surface charge distribution evolves due to two distinct mechanisms, namely  Ohmic currents from the bulk, $\llbracket \sigma {E}_n\rrbracket$, and advection by the particle surface velocity, $\bm{v}(\bm{x})$. Accordingly, the conservation equation for the surface charge is,
\begin{equation}
\partial_{t}q + \llbracket \sigma {E}_n\rrbracket +\bnabla_{s}\bcdot (q\bm{v})=0 \quad \mbox{for}\,\,\, \bm{x}\in S, \label{eq:chargeeq0}
\end{equation}
where $\bnabla_{s}\equiv ({\mathsfbi{I}}-\bm{nn})\bcdot\bnabla$ is the surface gradient operator. The fluid velocity field, $\bm{v}(\bm{x})$, and dynamic pressure,  $p(\bm{x})$, satisfy the Stokes equations in the suspending fluid, $-\mu \nabla^{2}\bm{v}+\bnabla p=\bm{0}$ and $\bnabla\bcdot\bm{v}=0$. No-slip at the solid-fluid interface leads to  kinematic boundary conditions for the fluid velocity, $\bm{v}(\bm{x}) = \bm{U} + \bm{\Omega} \btimes (\bm{x}-\bm{x}_c)$ for $\bm{x}\in S$, where $\bm{U}$, $\bm{\Omega}$ and $\bm{x}_c$ are the translational velocity, rotational velocity and centroid of the particle. In the absence of inertia, the dynamic balance of electric and hydrodynamic forces and torques on the solid particle requires $\bm{F}_E + \bm{F}_H = \bm{0}$ and $\bm{T}_E + \bm{T}_H = \bm{0}$, respectively. The forces and torques are found by integrating the surface tractions, $\bm{f}$,
\begin{align}
\bm{F}_{E,H} &= \oint_{S} \bm{f}_{E,H} \,\mathrm{d}S(\bm{x}), \label{eq:forceintegrate} \\
\bm{T}_{E,H} &= \oint_{S} (\bm{x}-\bm{x}_c) \btimes \bm{f}_{E,H} \,\mathrm{d}S(\bm{x}). \label{eq:torqueintegrate}
\end{align}
The electric and hydrodynamic tractions are expressed in terms of the Maxwell stress tensor, ${\mathsfbi{T}}_{E}$, and hydrodynamic stress tensor, ${\mathsfbi{T}}_{H}$, respectively as,
\begin{align}
\bm{f}_{E}&=\bm{n}\bcdot\mathsfbi{T}_{E}=\bm{n} \bcdot [\varepsilon (\bm{EE}-\tfrac{1}{2}E^{2}{\mathsfbi{I}})], \label{eq:electraction} \\
\bm{f}_{H}&=\bm{n}\bcdot\mathsfbi{T}_{H} = \bm{n}\bcdot[-p{\mathsfbi{I}}+\mu\left(\bnabla \bm{v}+\bnabla\bm{v}^{T}\right)]. \label{eq:stresses}
\end{align}
\begin{figure*}
    \centering
    \includegraphics[width=0.95\linewidth]{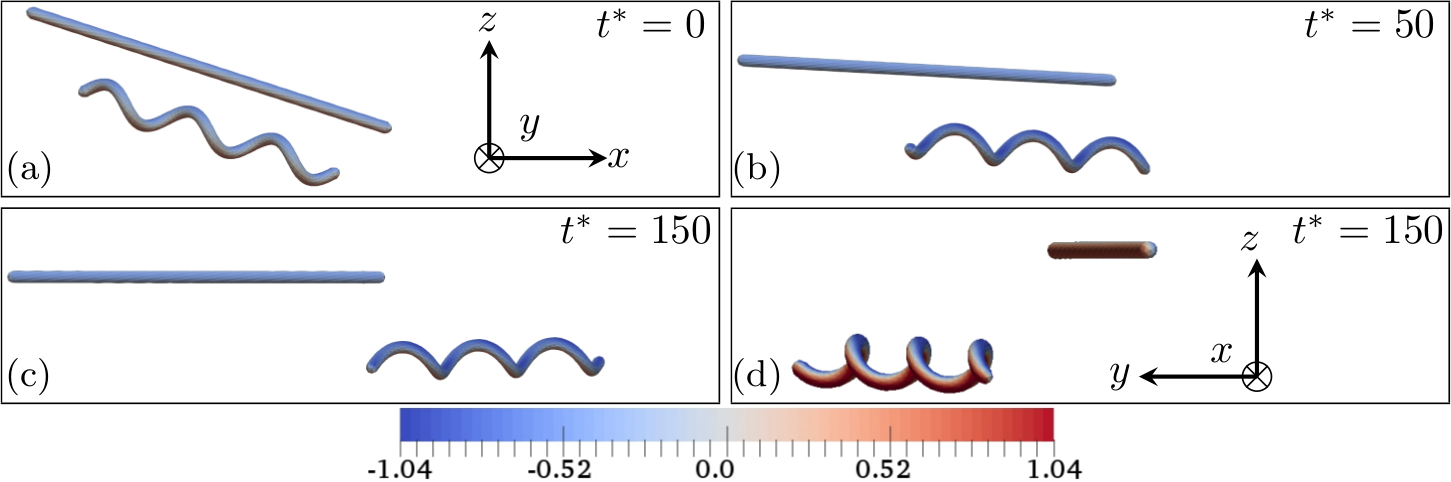}
    \caption{(Color online) (a--c) Snapshots of a cylinder and a helix having an aspect ratio of $a/L = 0.0167$ under Quincke rotation due to an applied electric field $\bm{E}_0^*=2.5 \bm{\hat{z}}$ with $R=Q=2$. The helix has $N=3$ turns, pitch angle $\alpha=0.2\pi$ and pitch $\lambda/L=0.236$. The particles are slightly tilted with respect to the $x$ axis at an angle $0.1\pi$ at time $t^*=0$. The cylinder performs pure rotation while the helix undergoes rotation as well as translation perpendicular to the $z$ axis. (d) Rotated view of snapshot (c) showing the positively charged side of the particles. The helix swims out of the $x-z$ plane due to its initially titled configuration. The colorbar indicates surface charge distribution. 
    }
    \label{fig:snapshots}
\end{figure*}
\nocite{fu2015} \nocite{pannacci2007} \nocite{pannacci2009} \nocite{lu2018}
To demonstrate that it is possible to convert Quincke rotation into spontaneous translation without the need for any surfaces, we consider a dielectric filament of helical shape in an infinite fluid. Helices are prototypical chiral particles used to create synthetic swimmers~\citep{zhang2009,ghosh2009} and their propulsive abilities at low Reynolds number flows have been well characterised in the context of bacterial locomotion~\citep{lauga16}. The centerline of the helix is specified as $r(\xi) = \xi \bm{\hat{x}} + R_h \cos{(2\pi \chi \xi/\lambda)} \bm{\hat{y}} + R_h \sin{(2\pi \chi \xi/\lambda)} \bm{\hat{z}}$ using parameter $\xi \in [-L_\lambda,L_\lambda]$, where $L_\lambda = N\lambda$ is the axial length, $\lambda$ is the helical pitch, $N$ is the number of turns and $R_h$ is the helical radius.  The arc and contour length of the helix are $s = \xi/\cos \alpha$ and $L=L_\lambda/\cos \alpha$, respectively, where $\alpha = \arctan (2\pi R_h/ \lambda)$ is the pitch angle. The cross-section of the helical filament is denoted as $a$. Here, $\chi=\pm 1$ determines the chirality of the helix and  we focus on right-handed helices, $\chi=1$, without any loss of generality. 

\begin{figure}
    \centering
    \includegraphics[width=0.95\linewidth]{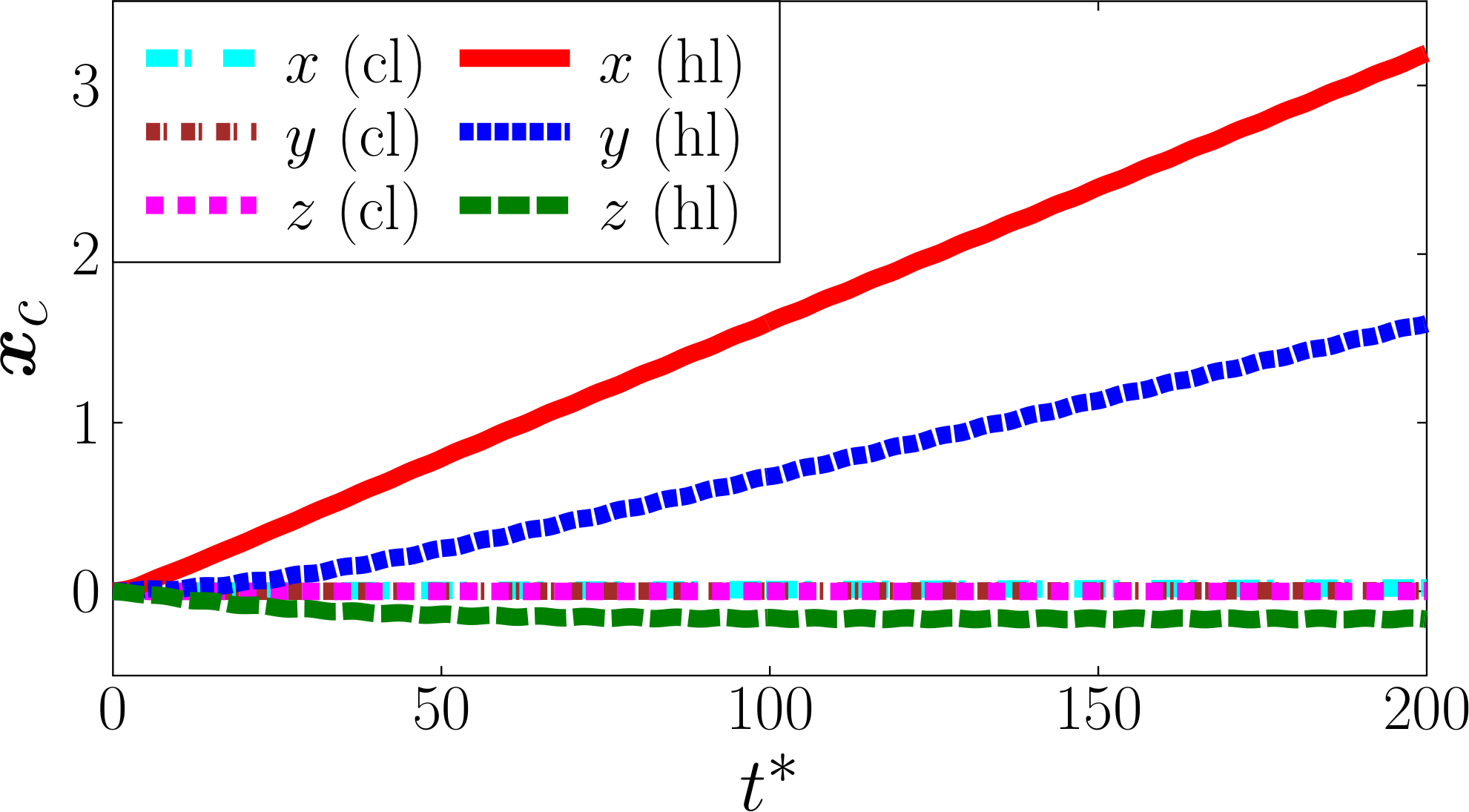}
    \caption{(Color online) Three-dimensional trajectories of the centroid of the cylinder (cl) and the helix (hl).}
    \label{fig:trajectory}
\end{figure}

We use the boundary element method to solve the electrohydrodynamics of a cylindrical and helical particle~\citep{pozrikidis2002,pozrikidis2011,das2017b} (see Supplemental Material~\citep{SM} for details). We show in Fig.~\ref{fig:snapshots}a-c snapshots of a cylinder and a helix having identical aspect ratio  (i.e.~the cylinder can be obtained by simply uncoiling the helix) moving under the action of an external uniform DC electric  field. We specify the dimensionless electric field strength, $E^* = E_0/E_{C,cl}$, where the critical electric field for Quincke rotation of a cylinder is $E_{C,cl} = \left({ {2\mu}/{\varepsilon^+ \tau_{\mathrm{MW},cl} (\overline{\varepsilon}_{cl} -\overline{\sigma}_{cl})}}\right)^{1/2}$ with $\overline{\varepsilon}_{cl}=(\varepsilon^- -\varepsilon^+)/((\varepsilon^- + \varepsilon^+)$ and $\overline{\sigma}_{cl}=(\sigma^- -\sigma^+)/(\sigma^- + \sigma^+)$~\citep{jones2005}. Time is non-dimensionalized with the characteristic Maxwell--Wagner timescale for polarization of a cylindrical particle upon the application of an electric field, $\tau_{\mathrm{MW},cl}=(\varepsilon^- +\varepsilon^+)/(\sigma^- + \sigma^+)$. The axes of both rigid particles are initially tilted at an angle of $0.1\pi$ with respect to the $x$ axis in the $x-z$ plane. Since the applied electric field, $\bm{E}_0^*=E_0^*\bm{\hat{z}}$, is higher than the critical field for both particles, they spontaneously start rotating. The directions of rotation for both particles are always perpendicular to the electric field, i.e. $\bm{\Omega}\bcdot \bm{E}_0 = 0$~\citep{SM}, and thus both align their axes in a direction perpendicular to the electric field in the steady state. As predicted by theory, the cylinder undergoes pure rotation with no translation. In contrast, the asymmetric shape of the  helix allows it to undergo both rotation and translation. Furthermore, we plot the net displacement of the cylinder (cl) and helix (hl) in three dimensions  in time, see Fig.~\ref{fig:trajectory}. Note that the helix swims out of the $x-z$ plane due to its initially tilted configuration.

\begin{figure*}[t]
    \centering
    \includegraphics[width=0.95\linewidth]{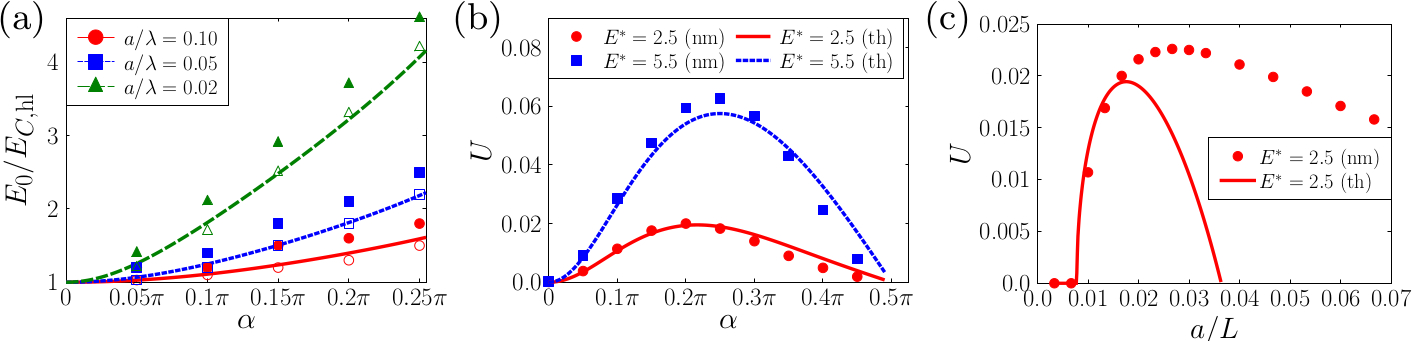}
        \caption{(Color online) (a) Critical electric field of a helix with three different cross-sectional radii but fixed number of turns, $N=1$, plotted against the pitch angle. Red, blue and green indicate $a/\lambda = 0.10,0.05,0.02$ respectively. Open and filled symbols indicate no-swimming and swimming, respectively, computed using numerical simulations based on boundary element method while solid and dashed lines represent $E^*/\sqrt{1+G}$ . (b) Swimming speed versus pitch angle for a helix with aspect ratio $a/L=0.0167$ for 2 different electric field strengths, $E^*=2.5$ (red circle and solid line indicate theory and simulations respectively) and $E^*=5.5$ (blue square and dashed line indicate theory and simulations respectively). (c) Swimming speed versus cross-sectional radius of a helix keeping the electric field $E^*=2.5$, number of turns $N=3$, pitch angle $\alpha=0.2\pi$ and pitch $\lambda/L=0.27$ fixed (red circle and solid line indicate theory and simulations respectively). 
        }
    \label{fig:3plots}
\end{figure*}

In contrast to Quincke rollers, the helical particle in Fig.~\ref{fig:snapshots} undergoes spontaneous translation in the absence of surfaces, and thus represents a new type of active self-propelling particle in bulk fluids. In order to further probe its ability to swim, we investigate in Fig.~\ref{fig:3plots} how its steady swimming speed, $U$, depends on various geometrical parameters (numerical data are shown in symbols while the lines represent the theory developed below). First we show in  Fig.~\ref{fig:3plots}a how the magnitude of the critical electric field depends on the  pitch angle, $\alpha$, for various cross-sectional radii, $a/\lambda$, with fixed number of turns. The critical field required to generate rotation of the helix is seen to systematically increase above its value for a cylinder as the amplitude of the helix grows and as the filament becomes more slender. 

Next we plot in Fig.~\ref{fig:3plots}b, the value of the steady swimming speed, $U$, as a function of the helix pitch angle, $\alpha$, for two different electric field strengths while keeping the cross-sectional radius fixed. The swimming speed is zero for a straight rod ($\alpha=0$) and a torus ($\alpha = \pi/2$) and thus is maximal when the  pitch angle takes an intermediate value, $\alpha\approx0.2\pi$ (simulations) and $0.215\pi$ (theory) for $E^*=2.5$ and, $\alpha\approx0.25\pi$ (simulations and theory) for $E^*=5.5$. Finally the effect of the aspect ratio of the helix, $a/L$, on the swimming speed, $U$, is shown in Fig.~\ref{fig:3plots}c keeping other geometrical quantities fixed. The swimming speed undergoes a supercritical pitchfork bifurcation so that swimming does not occur for $a/L$ below a critical value (i.e. for particles that are too slender). 

The computational results obtained above can be rationalised using theoretical arguments. The hydrodynamic forces and torques acting on a helix are linearly related to its translation and angular velocities through the $6\times 6$ resistance matrix $\mathsfbi{R}$ as,
\begin{equation}\label{eq:resistance}
  \begin{pmatrix}
	\bm{F}_H \\
	\bm{T}_H
  \end{pmatrix}
    =- \mathsfbi{R} \bcdot
    \begin{pmatrix}
	\bm{U} \\
	\bm{\Omega}
    \end{pmatrix}.
\end{equation}
The hydrodynamics of a helix can be described using the framework of resistive-force theory, which is valid for slender filaments moving in viscous fluids in the absence of inertia~\citep{lauga2009}. Assuming that the helix axis remains aligned with the $x$ direction, the components of the resistance matrix relevant for the analysis below are $\mathcal{R}_{44} = L R_h^2(\zeta_\perp \cos^2\alpha + \zeta_\parallel \sin^2\alpha)$, $\mathcal{R}_{11} = L(\zeta_\parallel \cos^2\alpha + \zeta_\perp \sin^2\alpha)$,  $\mathcal{R}_{14} = \chi LR_h^2\sin\alpha\cos\alpha  (\zeta_\parallel - \zeta_\perp)$ and $\mathcal{R}_{13}=\mathcal{R}_{34}=0$. All other elements of the resistance matrix are provided in the Supplemental Material~\citep{SM}. Here, $\zeta_\parallel$ and $\zeta_\perp$ are the drag coefficients for local motion of the helix along the directions parallel and perpendicular to its tangent~\citep{lighthill1976,SM}.  For the  electric problem, we assume that the helix is identical to a cylinder of the same contour length, a reasonable approximation if the  helix has a small pitch angle (i.e.~small amplitude).  The resulting electric and viscous torque acting on the helix are then given by,
\begin{align}
     \bm{T}_E & = 2\pi \varepsilon^+ a^2 L E_0^2 (\bm{P} \btimes \bm{E}_0), \label{eq:helixtorqueelec} \\ 
     \bm{T}_H &= -(4 \pi a^2 L \Omega_1 + \mathcal{R}_{44} \Omega_1 + \mathcal{R}_{14} U_1 )\bm{\hat{x}}, \label{eq:helixtorquehydro}
\end{align}
where $\bm{P}$ is the effective dipole moment of the helix. Since there is no electric force acting on the particle, we have $\bm{F}_E = -\bm{F}_H = \bm{0}$, leading to a relation between translational and angular velocity,
\begin{equation}\label{eq:swim}
U_1 = - \Omega_1(\mathcal{R}_{14}/\mathcal{R}_{11}).
\end{equation}
Balancing electric and viscous torques on the helix, $\bm{T}_E + \bm{T}_H = \bm{0}$, leads to  a relation between $P_2$ and $\Omega_1$,
\begin{equation}
    E^{*2}P_2/(\overline{\varepsilon}_{cl}-\overline{\sigma}_{cl}) - (1+G) \Omega_1  = 0, \label{eq:helixtorquebalance}
\end{equation}
where $G = (\mathcal{R}_{44}-\mathcal{R}_{14}^2/\mathcal{R}_{11})/(4\pi \mu a^2L)$ is a helical shape factor that only depends on geometry. The relaxation equation of the effective dipole moment of the helix derived from the charge conservation equation, Eq.~\eqref{eq:chargeeq0}, provides another relation between $P_2$ and $\Omega_1$~\citep{SM},
\begin{equation}
 P_2 = (\overline{\varepsilon}_{cl}-\overline{\sigma}_{cl})\Omega_1/(1+\Omega_1^2). \label{eq:helixdipole}
\end{equation} 
Eliminating $P_2$ from Eqs.~\eqref{eq:helixtorquebalance} and \eqref{eq:helixdipole}, we obtain two solutions for the angular velocity of a helix under Quincke rotation: (i) the trivial solution, $\Omega_1=0$, and (ii) the steady-state rotation solution,
\begin{equation}
    \Omega_1 = \sqrt{E^{*2}/(1+G) - 1}. \label{eq:helixangularvelocity}
\end{equation}
The critical electric field for Quincke rotation of a helix is then given as $E_{C,hl} = E_{C,cl} \sqrt{1+G}$ while the predicted swimming speed is given by Eq.~\eqref{eq:swim}. 

The predictions from this theoretical approach are compared with the computational results in Fig.~\ref{fig:3plots}. The theory is able to reproduce all features of the computational study, including the supercritical pitchfork bifurcation (at a fixed field strength) showing non-existence of swimming states for filaments that are too slender. This is because while the electric torque on the particle scales as $a^2$, the viscous torque scales as $a^2 + R_h^2$, see Eqs.~\eqref{eq:helixtorqueelec} and \eqref{eq:helixtorquehydro}. The breakdown of the theory for large values of $a/L$ is expected since the hydrodynamics based on resistive-force theory is accurate only in the asymptotic limit of slender filaments,  $a/L \rightarrow 0$.

In summary, we have shown in this Letter that the classical Quincke rotational instability of dielectric particles under DC electric fields can lead to spontaneous self-propulsion in a bulk fluid when combined with geometrical asymmetry. The phenomenon occurs  in the absence of any nearby surfaces, in stark contrast to Quincke rollers which require the presence of walls to break symmetries and swim. While a single particle rotates and translates in a plane perpendicular to the electric field, suspensions of such particles are expected to display out-of-plane swimming resulting from three-dimensional electrohydrodynamic interactions. As a practical example, we consider a helical particle made of Polymethyl methacrylate (PMMA) suspended in various classical dielectric fluids and predict swimming speeds of tens of microns per second (see Supplemental Material~\citep{SM}). The physical mechanism of this new form of self-propulsion was  demonstrated using numerical computations for the full system in the case of a helical filament and confirmed analytically by a theoretical approach in the slender-helix limit. Though we have focused on the special case of helical particles, self-propulsion is expected to occur for any kind of asymmetric particles whose resistance matrix, $\mathsfbi{R}$, contains a nonzero off-diagonal term enabling coupling of an imposed rotation to translation.  
Suspensions of randomly-shaped particles under Quincke rotation interacting electrohydrodynamically are thus expected to perform collective motion by exploring the full three-dimensional space, thereby, opening doors to a potentially new type of active matter.

\smallskip

We  thank David Saintillan for helpful discussions. This project has received funding from the European Research Council (ERC) under the European Union's Horizon 2020 research and innovation programme (grant agreement 682754 to EL).

\bibliography{papers} 
\bibliographystyle{apsrev4-1}

\newpage

\appendix 
\onecolumngrid
\section{Supplemental Material}
\subsection{Numerical Method}\label{sec:numerics}

The electrohydrodynamics of a dielectric particle, governed by Laplace and Stokes equations, is best solved using the boundary element method~\citep{pozrikidis2002,pozrikidis2011}. The electric potential is represented in terms of the single-layer density $\llbracket {E}_{n}(\bm{x})\rrbracket$ as,
\begin{equation}
    \varphi(\bm{x}_{0}) =-\bm{x}_0\bcdot \bm{E}_{0}+\oint_{S} \llbracket {E}_{n}(\bm{x})\rrbracket \, \mathcal{G}(\bm{x}_0;\bm{x})\,\mathrm{d}S(\bm{x}), \label{eq:intpotential}
\end{equation}
where $\bm{x}_{0}\in V^{\pm}, S$, $\bm{x} \in S$ and the Green's function or fundamental solution of Laplace's equation in an unbounded domain is given by $\mathcal{G}(\bm{x}_{0};\bm{x})=1/{4\pi r}$, $\bm{r}=\bm{x}_{0}-\bm{x}$ and  $r=|\bm{r}|$. For a given surface charge distribution $q(\bm{x})$ at any time, we first compute the jump in normal electric field across the interface $\llbracket {E}_{n}\rrbracket$ using an integral equation derived from manipulating Eq.~\eqref{eq:intpotential},
\begin{align}
\begin{split}
&\oint_{S} \{\llbracket {E}_{n}(\boldsymbol{x})\rrbracket-\llbracket {E}_{n}(\boldsymbol{x}_0)\rrbracket\}\{\boldsymbol{n}(\boldsymbol{x}_0)\boldsymbol{\cdot}\bnabla_{0}\mathcal{G}(\boldsymbol{x}_0;\boldsymbol{x})\}\,\mathrm{d}S(\boldsymbol{x})\\
& +\llbracket {E}_{n}(\boldsymbol{x}_0)\rrbracket \left[\frac{Q}{Q-1}-L(\boldsymbol{x}_0)  \right]=E_{n0}+\frac{q(\boldsymbol{x}_0)}{Q-1},
\label{eq:intjump}
\end{split}
\end{align}
where $\boldsymbol{x}_{0}\in S$ and $L$ is a purely geometric quantity~\citep{das2017b}. Having computed $\llbracket {E}_{n}\rrbracket$, we can use integral Eq.~\eqref{eq:intpotential} to compute the electric potential. The tangential component of the electric field, $\bm{E}_t$, is computed numerically by taking tangential derivatives of the electric potential. The normal components of the electric field are easily obtained using Gauss's law,
\refstepcounter{equation}
$$
E_n^+=\frac{q-Q\llbracket {E}_{n}\rrbracket}{1-Q}, \qquad {E}_n^-=\frac{q-\llbracket {E}_{n}\rrbracket}{1-Q}. \eqno{(\theequation{\mathit{a},\mathit{b}})} \label{eq:normalelectric}
$$
Finally, we determine the jump in the normal component of Ohmic currents $\llbracket \sigma E_n\rrbracket$ and the external electric traction $\bm{f}_E$ using Eq.~(4) in the main text. The net electric force and torque acting on the particle is found by integrating electric traction using Eqs.~(2) and (3) in the main text. The next step involves computing the hydrodynamic force and torque by using the dimensionless form of force and torque balance equations, i.e. $\bm{F}_H = -\bm{F}_E/Ma$ and $\bm{T}_H = -\bm{T}_E/Ma$. Here, we have introduced the third dimensionless Mason number, $Ma$, ($R=\sigma^+/\sigma^-$ and $Q= \varepsilon^-/\varepsilon^+$ being the other two) that denotes the ratio of viscous to electric stresses,
\begin{equation}
    Ma = \frac{\mu}{\varepsilon^+ \tau_{\mathrm{MW}}E_{0}^2}.
\end{equation}
In this work, we have chosen to specify the dimensionless electric field strength $E^* = E_0/E_{C}$ instead of the Mason number, $Ma$, where the critical electric field for Quincke rotation of a particle is,
\begin{equation}
E_{C} = \sqrt{\frac{2\mu}{\varepsilon^+ \tau_{\mathrm{MW}} (\overline{\varepsilon} -\overline{\sigma})}}.
\end{equation}
These two dimensionless numbers are related as $Ma =(\overline{\varepsilon}- \overline{\sigma})/(2 E^{*2})$. The only term remaining to be computed in the charge conservation equation is the surface velocity $\bm{v}(\bm{x} \in S)$. This is obtained by solving the hydrodynamic problem for the particle subject to a force and a torque. Assuming creeping flow, we use the Stokes boundary integral equation to represent the particle surface velocity as,
\begin{equation}
\bm{U} + \bm{\Omega} \btimes (\bm{x}_0-\bm{x}_c)  = -\frac{1}{8\pi\mu}\oint_S \bm{f}_h (\bm{x}) \bm{\cdot} \mathsfbi{G}(\bm{x}_0;\bm{x}) \,\mathrm{d}S(\bm{x}), \label{eq:stokesbie}
\end{equation}
where $\bm{x}_0 \in S$ and $\mathsfbi{G}(\bm{x}_0;\bm{x})$ denotes the free-space Green's functions for the Stokeslet $\mathsfbi{G}(\bm{x}_0;\bm{x})=\mathsfbi{I}/r + \bm{r}\bm{r}/r^3$.  The Stokes boundary integral equation \eqref{eq:stokesbie} and the equations relating the net force and torque to the surface tractions, Eqs.~(2) and (3) in the main text, are solved together to find the unknown particle velocities $\bm{U}$ and $\bm{\Omega}$. Having computed the surface velocity, the final step involves numerically integrating the charge conservation equation (without the charge convection part since the nodes are advected with the interfacial velocity) in time using a second order Runge-Kutta scheme until a steady state is reached.
 
\section{Validations: Quincke rotation of a sphere and a cylinder}

 \begin{figure*}
     \centering
     \includegraphics[width=0.95\linewidth]{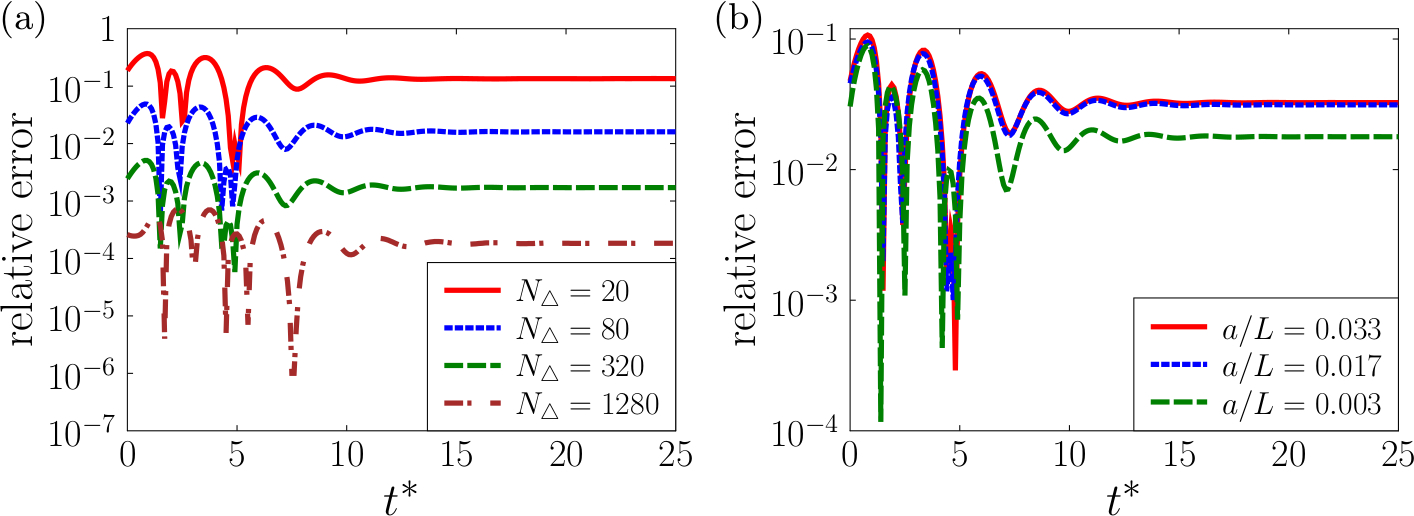}
         \caption{(Color online) (a) Relative error in the angular velocity of a sphere under Quincke rotation computed using boundary element method for various grid sizes, where $N_\triangle$ is the total number of elements. (b) Relative error in the angular velocity of a cylinder under Quincke rotation computed using boundary element method for various aspect ratios $a/L$ but fixed number of elements, $N_\triangle=444$.}
     \label{fig:angular}
 \end{figure*}
 
Using spherical harmonics for a sphere (sp) under Quincke rotation, we can obtain the dipole moment, $\bm{P}$, relaxation equation from the charge conservation equation (Eq.~(1) in the main article),
 \begin{equation}
     \frac{d \bm{P}}{dt} = \bm{\Omega} \btimes [\bm{P} - a^3 \overline{\varepsilon}_{sp}\bm{E_0}] -\frac{1}{\tau_{\mathrm{MW},sp}} [\bm{P} - a^3 \overline{\sigma}_{sp}\bm{E_0}], \label{eq:dipolerelax}
 \end{equation}
 where $a$ is the radius of the sphere, $\tau_{\mathrm{MW},sp}=(\varepsilon^- +2\varepsilon^+)/(\sigma^- + 2\sigma^+)$ is the characteristic timescale for polarization of the spherical particle upon application of the field. The other two dimensionless numbers, $\overline{\varepsilon}_{sp}=(\varepsilon^- -\varepsilon^+)/(\varepsilon^- + 2\varepsilon^+)$ and $\overline{\sigma}_{sp}=(\sigma^- -\sigma^+)/(\sigma^- + 2\sigma^+)$ are called the Clausius--Mossotti factors. It can be easily shown that quadrupoles and higher multipoles are absent if the electric field is uniform and these multipoles do not interact with each other \citep{das2013}. As there is no net force on the particle, $\bm{F}_E = - \bm{F}_H= \bm{0}$, we only need the torque balance equation to determine the angular velocity of the spherical particle,
 \begin{equation}
     4 \pi \varepsilon^+ \bm{P} \btimes \bm{E}_0 - 8\mu a^3 \bm{\Omega} = \bm{0}. \label{eq:torquebalancesphere}
\end{equation}
 The dipole moment scales as $a^3 E_0$ and the angular velocity scales as $\tau_{\mathrm{MW},sp}^{-1}$. If the applied field strength $E_0$ exceeds the critical electric value $E_{C,sp}$, the steady-state angular velocity of a sphere under Quincke rotation is,
\begin{align}
     &\Omega=\frac{1}{\tau_{\mathrm{MW},sp}}\sqrt{\frac{E_0^2}{E_{C,sp}} - 1}, \quad \mbox{where}\\  \label{eq:omegasphere}
     & E_{C,sp} = \sqrt{\frac{2\mu}{\varepsilon^+ \tau_{\mathrm{MW},sp} (\overline{\varepsilon}_{sp} - \overline{\sigma}_{sp})}}.
\end{align}
We can perform the same analysis for an infinitely long cylinder (cl) of cross-sectional radius $a$ using polar harmonics. The dipole moment relaxation equation \eqref{eq:dipolerelax} retains the same form, except for two changes, namely, (i) $a^3$ is replaced with $a^2$, and (ii) the Clausius--Mossotti factors and the Maxwell--Wagner relaxation times change to $\overline{\varepsilon}_{cl}=(\varepsilon^- -\varepsilon^+)/((\varepsilon^- + \varepsilon^+)$, $\overline{\sigma}_{cl}=(\sigma^- -\sigma^+)/(\sigma^- + \sigma^+)$ and $\tau_{\mathrm{MW},cl}=(\varepsilon^- +\varepsilon^+)/(\sigma^- + \sigma^+)$, respectively. These changes are a direct consequence of the form of the electric potential in polar harmonics, $\phi^+ =  \bm{P}\bcdot \bm{x}/ x^2$, as compared to that in spherical harmonics, $\phi^+ =  \bm{P}\bcdot \bm{x}/ x^3$. The torque balance equation per unit length for a cylinder is given as,
\begin{equation}
     2 \pi \varepsilon^+ \bm{P} \btimes \bm{E}_0 - 4\mu a^2 \bm{\Omega} = \bm{0}. \label{eq:torquebalancecyl}
 \end{equation}
The steady-state angular velocity of an infinitely long cylinder under Quincke rotation is,
\begin{align}
     &\Omega=\frac{1}{\tau_{\mathrm{MW},cl}}\sqrt{\frac{E_0^2}{E_{C,cl}} - 1}, \quad \mbox{where}\\ \label{eq:omegacylinder}
     & E_{C,cl} = \sqrt{\frac{2\mu}{\varepsilon^+ \tau_{\mathrm{MW},cl} (\overline{\varepsilon}_{cl} - \overline{\sigma}_{cl})}}.
\end{align}

The 2 coupled ODEs Eqs.~\eqref{eq:dipolerelax} and \eqref{eq:torquebalancesphere}, relevant for a sphere or Eqs.~\eqref{eq:dipolerelax} (with appropriate changes) and \eqref{eq:torquebalancecyl}, relevant for a cylinder can be marched in time with a specified initial condition and serve as a comparison for the numerical results. For a given dipole moment $\bm{P} (t=0) = \bm{P}_0$, we can find the surface charge distribution using Gauss's Law $q(\bm{x},t=0)=q_0$ and solve the relevant integral equations described in \S~I. Simultaneously, we can also solve the ODEs to find the angular velocity at any given time. In Fig.~\ref{fig:angular}(a), we plot the relative error in the angular velocity of a sphere under Quincke rotation obtained by numerical simulations for various grid sizes. We find that the numerical results converge to the theoretical one as the grid size is decreased. In Fig.~\ref{fig:angular}(b), we plot the relative error in the angular velocity of a cylinder having various aspect ratios while the total number of elements is kept fixed, $N_\triangle=444$. We find that the agreement between theory and numerics is the best for the  cylinder with the lowest aspect ratio. This is expected as the theory is valid for an infinite long cylinder, $a/L \rightarrow 0$. 

\section{Resistance Matrix}
For the sake of completeness, we provide all the elements of the resistance matrix $\mathsfbi{R}$ relevant for resistive-force theory of a slender helix in viscous flow \citep{fu2015},
\begin{equation}
    \begin{bmatrix}
        F_{h,1} \\
        F_{h,2} \\
        F_{h,3} \\
        T_{h,1} \\
        T_{h,2} \\
        T_{h,3} \\
    \end{bmatrix} = 
    -\begin{bmatrix}
        \mathcal{R}_{11} & 0 & \mathcal{R}_{13} & \mathcal{R}_{14} & 0 & \mathcal{R}_{16} \\
        0 & \mathcal{R}_{22} & 0 & 0 & \mathcal{R}_{25} & 0 \\
        \mathcal{R}_{13} & 0 & \mathcal{R}_{33} & \mathcal{R}_{34} & 0 & \mathcal{R}_{36} \\
        \mathcal{R}_{14} & 0 & \mathcal{R}_{34} & \mathcal{R}_{44} & 0 & \mathcal{R}_{46} \\
        0 & \mathcal{R}_{25} & 0 & 0 & \mathcal{R}_{55} & 0 \\
        \mathcal{R}_{16} & 0 & \mathcal{R}_{36} & \mathcal{R}_{46} & 0 & \mathcal{R}_{66} 
    \end{bmatrix}
    \begin{bmatrix}
        U_1 \\
        U_2 \\
        U_3 \\
        \Omega_1 \\
        \Omega_2 \\
        \Omega_3 \\
    \end{bmatrix}.
\end{equation}
where,
\begin{subequations}
\begin{eqnarray}
&\mathcal{R}_{11} &= L(\zeta_\parallel \cos^2\alpha + \zeta_\perp \sin^2\alpha), \\
&\mathcal{R}_{13} &= 0, \\
&\mathcal{R}_{14} &= \chi LR_h \sin\alpha \cos\alpha (\zeta_\parallel - \zeta_\perp), \\
&\mathcal{R}_{16} &= LR_h \sin\alpha \cos\alpha (\zeta_\parallel - \zeta_\perp), \\
&\mathcal{R}_{22} &= \tfrac{1}{2}L [\zeta_\perp(1+\cos^2\alpha) + \zeta_\parallel \sin^2\alpha ], \\
&\mathcal{R}_{25} &= -\tfrac{3}{4}\mathcal{R}_{14}, \\
&\mathcal{R}_{33} &= \mathcal{R}_{22}, \\
&\mathcal{R}_{34} &= 0, \\
&\mathcal{R}_{36} &= -\tfrac{1}{4}\mathcal{R}_{14}, \\
&\mathcal{R}_{44} &= L R_h^2(\zeta_\perp \cos^2\alpha + \zeta_\parallel \sin^2\alpha), \\
&\mathcal{R}_{55} &= \tfrac{1}{12}[\zeta_\perp L\{ L^2\cos^4\alpha + 6R_h^2 \sin^2\alpha \} + LR_h^2\cos^2\alpha \{(2N^2\pi^2 + 15)\zeta_\parallel  \nonumber\\
&+& (2N^2\pi^2 - 9)\zeta_\perp \}] \\
&\mathcal{R}_{66} &= \tfrac{1}{12}[\zeta_\perp L\{ L^2\cos^4\alpha + 6R^2 \sin^2\alpha \} 
+ LR_h^2\cos^2\alpha \{(2N^2\pi^2 - 3)\zeta_\parallel \nonumber\\
&+& (2N^2\pi^2 + 9)\zeta_\perp \}].
\end{eqnarray}
\end{subequations}
The drag coefficients along the directions parallel and perpendicular to the tangent of the centerline representing the slender particle are,
\refstepcounter{equation}
$$
    \zeta_\parallel = \frac{2\pi\mu}{\ln\left( {0.18\lambda}/{a\cos \alpha}\right)}, \qquad \zeta_\perp = \frac{4\pi\mu}{\ln \left( {0.18\lambda}/{a\cos \alpha}\right) +0.5}.
   \eqno{(\theequation{\mathit{a},\mathit{b}})} 
 \label{eq:RFT} 
$$ 

\section{Dimensional values of the swimming speed}

In this section, we discuss the order of magnitude of the dimensional values of (a) the critical electric field for various solid-liquid systems, and (b) the swimming speed achievable by a helical particle under Quincke rotation. Using data from past literature, a dielectric helical particle can be made using Poly-methyl-methacrylate (PMMA) and a dielectric liquid \citep{pannacci2007,pannacci2009,bricard2013,bricard2015}, see Table \ref{tbl:forexpt}. The permittivity, conductivity, and density of PMMA are $\varepsilon^- = 2.3 \varepsilon_0$ and $\sigma^- = 10^{-14}$ \si{\siemens\per\metre}, and $\rho^- = 1.18\times10^3$ \si{\kilogram\per\milli\metre\cubed}, respectively, where $\varepsilon_0 = 8.8542 \times 10^{-12}$ \si{\farad\per\metre} is the permittivity of vacuum.  When performing experiments, the density of the particle and suspending liquid must be matched by adding suitable agents to the liquid.
\begin{table}[t]
\centering
\begin{tabularx}{\textwidth}{ |c| *{10}{Y|} }
\cline{2-7}
\hline
 Liquid & Permittivity & Conductivity & Viscosity & Density & MW time & Critical Field \\ \hline
  & $\varepsilon^+$ & $\sigma^+$ (\si{\nano\siemens\per\metre}) & $\mu$ (\si{\milli \pascal\second}) & $\rho^+$ (\si{\kilogram\per\metre\cubed}) & $\tau_{\mathrm{MW},sp|cl}$  (\si{\milli\second}) & $E_{C,sp|cl}$ (\si{\volt\per\micro\metre})\\ \hline
 Dodecane & 2.17 $\varepsilon_0$ & 50 & 1.64 & 770 & $0.6|0.9$ & $0.703|0.430$ \\ \hline
 Dielec S & 2.4 $\varepsilon_0$ & 4.3 & 12.9 & 840 & $7.6|10.3$ & $0.550|0.337$ \\ \hline
  Dielec S + Ugilec & 3.69 $\varepsilon_0$ & 33 & 13.6 & 1180 & $1.3|1.7$ & $1.261|0.773$ \\ \hline
  Hexadecane & 2.2 $\varepsilon_0$ & 140  & 3 & 770 & $0.22|0.30$ & $1.580|0.968$ \\ \hline
\end{tabularx}
    \caption{Physico-chemical properties of various dielectric liquids, the Maxwell-Wagner relaxation time and critical electric field of a solid spherical or cylinder particle made of PMMA under Quincke rotation in these liquids.}
\label{tbl:forexpt}
\end{table}

Let us consider a helical particle made of PMMA material suspended in a hexadecane solution and subject to an electric field of strength $E=2.5E_{C,cl} = $ \SI{2.42}{\volt\per\micro\metre}. These magnitudes of electric field strength have been employed in recent experiments involving Quincke rotation \citep{lu2018}. The helical parameters are chosen as, contour length $L=$ \SI{3}{\micro\meter}, $N=3$ turns, pitch angle $\alpha=0.2\pi$, and radius of cross-section $a=$ \SI{0.08}{\micro\meter} and \SI{0.16}{\micro\meter}. Using all the relevant data, the numerical simulations predict a swimming speed of $U=$ \SI{75}{\micro\meter\per\second} and \SI{61}{\micro\meter\per\second}.  
It is noteworthy that for a given helical shape and fixed critical electric field strength, the dimensional swimming speed value is inversely proportional to the Maxwell-Wagner relaxation time. The diffusivity of these micron-sized particle has not been taken into account in this work. It is typically of the order of $D^{-1}=0.31$ \si{\second} corresponding to a persistence length, $l_p \sim$ \SI{25}{\micro\meter} \citep{bricard2013}.
\end{document}